\documentclass[aps,showpacs,preprintnumbers]{revtex4}
\usepackage{amssymb}
\usepackage{amsmath}
\usepackage{bm}

\input{tcilatex}

\begin{document}

\title{Antiparticle Contribution in the Cross Ladder Diagram for
Bethe-Salpater Equation in the Light-front.}
\author{J.H.O. Sales$^{1,2}$}
\affiliation{$^{1}$Funda\c{c}\~{a}o de Ensino e Pesquisa de Itajub\'{a}, CEP 37501-002,
Itajub\'{a}, MG, Brazil.\\
$^{2}$Instituto de Ci\^{e}ncias Exatas, Universidade Federal de Itajub\'{a},
CEP 37500-000, Itajub\'{a}, MG, Brazil}
\author{A.T. Suzuki}
\affiliation{Instituto de F\'{\i}sica Te\'{o}rica/UNESP, Rua Pamplona, 145 CEP 01405-900 S%
\~{a}o Paulo, SP, Brazil}
\date{\today}

\begin{abstract}
We construct the homogeneous integral equation for the vertex of the bound
state in the light front with the kernel approximated to order $g^{4}$. We
will truncate the hierarchical equations from Green functions to construct
dynamical equations for the two boson bound state exchanging interacting
intermediate $\sigma $ bosons and including pair creation process
contributing to the crossed ladder diagram.

.
\end{abstract}

\pacs{12.39.Ki,14.40.Cs,13.40.Gp}
\maketitle

\address{ 27607.}









\section{Introduction}

In relativistic quantum field theories the Bethe-Salpeter equation describes
a system of two interacting particles \cite{37}. In general, for practical
applications, to study the bound state of two particles, its kernel is
truncated in its lowest order. Today, it is already possible to solve the
Bethe-Salpeter equation including the \textquotedblleft
ladder\textquotedblright\ diagram and \textquotedblleft crossed
ladder\textquotedblright\ diagrams in its kernel.\cite{karmanov}. Also,
recently, the Bethe-Salpeter equation in the \textquotedblleft
ladder\textquotedblright\ approximation has been solved in Minkowski space 
\cite{carbonel}. However up to the time of the publication of J.H.O.Sales
work \cite{13,14} there was no detailed comparison between the results in
the four dimensional calculations and in the light front, with the kernel
including the propagation of intermediate states of four bodies. The
objective was to study the effect of the components of the Fock space with
particle number greater than three, going beyond the previously done works
in the area.

The concentrated effort to use the tridimensional reduction in light front
coordinates is justifiable, since there is hope that using light front
quantization would make it possible to better understand several aspects of
low energy quantum chromodynamics (QCD) \cite{36}. In the past, the concept
of wave function in the light front was applied in the context of nuclear
physics to describe the deuteron \cite{30,5} and the discussion of its
properties in the light front extends up to now \cite{32}.

The representation of the Bethe-Salpeter equation in the infinite momentum
frame was studied by Weinberg \cite{10}, which corresponds to truncation of
the Fock space in the light front up to the intermediate state of three
particles. In this approximation numerical results in different contexts
have been obtained, such as in bosonic models \cite{11} and in fermionic
ones \cite{17}. However, an explicit systematic expansion for the
Bethe-Salpeter equation in the light front was still lacking in the
literature. Although in the past there had been attempts to include
intermediate states of more particles but in an incomplete way \cite{41}.

In this work we explore the concept of covariant quantum propagator written
down in terms of light front coordinates and obtain the propagator and
Green's function in the light-front for a time interval $x^{+}$ where $%
x^{+}=t+z$, is the light-front \textquotedblleft time\textquotedblright . In
principle, this is equivalent to the canonical quantization in the light
front \cite{20,36}. Kogut and Soper \cite{22} also makes use of this way of
constructing quantities in the light front: starting with $4$-dimensional
amplitudes or equations, they integrate over $k^{-}=k^{0}-k^{3}$, which
plays the role of \textquotedblleft energy\textquotedblright\ and
corresponds to processes described by amplitudes or equations in
\textquotedblleft time\textquotedblright\ $x^{+}$. With this, the relative
time between particles disappear and only the global propagation of
intermediate state is allowed. The global propagation of the intermediate
state is the \textquotedblleft time\textquotedblright\ translation of the
physical system between two instants $x^{+}$ and $x^{^{\prime }+}$.

\section{Free boson}

The propagation of a free particle with spin zero in four dimensional
space-time is represented by the covariant Feynman propagator 
\begin{equation}
S(x^{\mu })=\int \frac{d^{4}k}{\left( 2\pi \right) ^{4}}\frac{ie^{-ik^{\mu
}x_{\mu }}}{k^{2}-m^{2}+i\varepsilon },  \label{2.1}
\end{equation}%
where the coordinate $x^{0}$ represents the time and $k^{0}$ the energy. We
are going to calculate this propagation in the light front, that is, for
times $x^{+}$.

A point in the four-dimensional space-time is defined by the set of numbers $%
(x^{0},x^{1},x^{2},x^{3})$, where $x^{0}$ is the time coordinate and $%
\mathbf{x}=(x^{1},x^{2},x^{3})$ is the three-dimensional vector with space
coordinates $x^{1}=x$, $x^{2}=y$ e $x^{3}=z$. Observe that we adopt here the
usual convention to take the speed of light as $c=1$.

In the light front, time and space coordinates are mixed up and we define
the new coordinates as follows: 
\begin{equation}
x^{+}=x^{0}+x^{3},\text{ }x^{-}=x^{0}-x^{3}\text{ \ e \ }\vec{x}^{\perp
}=x^{1}\overrightarrow{i}+x^{2}\overrightarrow{j},
\end{equation}%
where $\overrightarrow{i}$ and $\overrightarrow{j}$ are unit vectors in the
direction of $x$ and $y$ coordinates respectively.

The null plane is defined by $x^{+}=0$, that is, this condition defines a
hyperplane that is tangent to the light-cone, the reason why many authors
call the hypersurface simply by light-cone.

The initial boundary conditions for the dynamics in the light front are
defined on this hyperplane. The axis $x^{+}$ is perpendicular to the plane $%
x^{+}=0$. Therefore a displacement of such hyperplane for $x^{+}>0$ is
analogous to the displacement of a plane in $t=0$ to $t>0$ of the
four-dimensional space-time. With this analogy, we recognize $x^{+}$ as the
time in the null plane.

We make the projection of the propagator for a boson in time associated to
the null plane rewriting the coordinates in terms of time coordinate $x^{+}$
and the position coordinates $(x^{-}$ and $\vec{x} _{\perp })$. With these,
the momenta are given by $k^{-}$, $k^{+}$ and $\vec{k}_{\perp }$, and
therefore we have 
\begin{equation}
S(x^{+})=\frac{1}{2}\int \frac{dk_{1}^{-}}{\left( 2\pi \right) }\frac{ie^{%
\frac{-i}{2}k_{1}^{-}x^{+}}}{k_{1}^{+}\left( k_{1}^{-}-\frac{k_{1\perp
}^{2}+m^{2}-i\varepsilon }{k_{1}^{+}}\right) }.  \label{2.2}
\end{equation}

The Jacobian of the transformation $k^{0}$, $\vec{k}\rightarrow k^{-},k^{+},%
\vec{k}_{\perp }$ is equal to $\frac{1}{2}$ and $k^{+} $, $k_{\bot }$ are
momentum operators.

Evaluating the Fourier transform, we obtain 
\begin{equation}
\widetilde{S}(k^{-})=\int dx^{+}e^{\frac{i}{2}k^{-}x^{+}}S(x^{+}),
\label{2.3}
\end{equation}%
where we have used 
\begin{equation}
\delta (\frac{k^{-}-k_{1}^{-}}{2})=\frac{1}{2\pi }\int dx^{+}e^{\frac{i}{2}%
\left( k^{-}-k_{1}^{-}\right) x^{+}},  \label{2.4}
\end{equation}%
and the property of Dirac's delta \textquotedblleft
function\textquotedblright\ $\delta \left( ax\right) =\frac{1}{a}\delta
\left( x\right) $ and we get 
\begin{equation}
\widetilde{S}(k^{-})=\frac{i}{k^{+}\left( k^{-}-\frac{k_{\perp
}^{2}+m^{2}-i\varepsilon }{k^{+}}\right) },  \label{2.6}
\end{equation}%
which describes the propagation of a particle forward to the future and of
an antiparticle backwards to the past. This can be oberved by the
denominator which hints us that for $x^{+}>0$ and $k^{+}>0$ we have the
particle propagating forward in time of the null plane. On the other hand,
for $x^{+}<0$ and $k^{+}<0$ we have an antiparticle propagating backwards in
time.

The Green funtion in the light front $G(x^{+})$ acting in the Fock space is
defined as the probability amplitude of the transition from the initial
state in the Fock space $\left| i\right\rangle $ to the final state $\left|
f\right\rangle $. Its Fourier transform is sometimes called resolvent for a
given Hamiltonian \cite{18}, however, here we call simply Fourier transform
for the Green function or even Green function itself.

In the case of a free boson, the Green function for the propagation of a
particle is defined by the operator 
\begin{equation}
G_{0}^{(1p)}(k^{-})=\frac{\theta (k^{+})}{k^{-}-k_{on}^{-}+i\varepsilon }\ ;
\label{2.7}
\end{equation}
where $k_{on}^{-}=\frac{k_{\perp }^{2}+m^{2}}{k^{+}}$ is the energy of the
particle. For the antiparticle propagation, we have: 
\begin{equation}
G_{0}^{(1a)}(k^{-})=\frac{\theta (-k^{+})}{k^{-}-k_{on}^{-}-i\varepsilon }\ .
\label{2.8}
\end{equation}

We can see that the difference between the Green functions in (\ref{2.7})
and (\ref{2.8}) for the propagator in the light front is the absence of the
imaginary (complex) number $i$ and of the factor of phase space $k^{+}$
which appears in (\ref{2.6}).

The operator defined by (\ref{2.7}) is the Green function of 
\begin{equation}
\left( k^{-}-k_{on}^{-}\right) \left(
G_{0}^{(1p)}(k^{-})+G_{0}^{(1a)}(k^{-})\right) =1\ .  \label{2.9}
\end{equation}

The Feynman propagator is then rewritten as: 
\begin{equation}
S(k^{\mu })=\frac{i}{k^{+}}G_{0}^{(1p)}(k^{-})-\frac{i}{|k^{+}|}%
G_{0}^{(1a)}(k^{-})=\frac{i}{k^{+}(k^{-}-\frac{k_{\perp
}^{2}+m^{2}-i\varepsilon }{k^{+}})}\ .  \label{2.10}
\end{equation}

\section{Green function of two bosons}

Our aim in this chapter is to study the two body Green function in the
``ladder'' approximation for the dynamics defined in the light front. Within
this treatment, we are not going to deal with perturbative corrections that
can be decomposed into one body problem.

Our interest is to define in the light front, the interaction between two
bodies mediated by the interchange of a particle and obtain the correction
to the two body Green function originated in this interaction.

For this purpose, we use a bosonic model for which the interaction
Lagrangian is defined as: 
\begin{equation}
\mathcal{L}_{I}=g\phi _{1}^{\ast }\phi _{1}\sigma +g\phi _{2}^{\ast }\phi
_{2}\sigma ,  \label{lagr}
\end{equation}%
where the bosons $\phi _{1}$ and $\phi _{2}$ have equal mass $m$ and the
intermediate boson, $\sigma $, has the mass $m_{\sigma }$. The coupling
constant is $g$.

Taking from Dirac's idea \cite{1} of representing the dynamics of a quantum
system in the light front in time $x^{+}=t+z$, we derive in this chapter the
two body Green function or covariant propagator which describes the
evolution of the system from a hypersurface $x^{+}=\mathrm{constant}$ to
another one. The Green function in the light front is the probability
amplitude for a initial state in $x^{+}=0$ evolving to a final state in $%
x^{+}>0$, where the evolution operator is defined by the Hamiltonian in the
light front \cite{2}. Sometimes we call the resolvent $\left( Z-H\right)
^{-1}$ as the Green function, too \cite{36}.

The two body Green function in the light front includes the propagation of
intermediate states with any number of particles.

We start our discussion evaluating the second order correction to the
coupling constant associated with the propagator. We define the matrix
element for the interaction and so we obtain the correction to the Green
function in the light front. Then we evaluate the correction to the Green
function to the fourth order in the coupling constant, where we use the
technique of factorizing the energy denominators, which is important to
identify the global propagation of four bodies after the integration in the
energies $k^{-}$. We discuss the generalization of this technique.

We show how to get the non perturbative Green function from the set of
hierarchical equations for the Green function in the \textquotedblleft
ladder\textquotedblright\ approximation. This corresponds to the truncation
of the Fock space in the light front, such that, only states with two bosons 
$\phi _{1}$ and $\phi _{2}$ are allowed, with no restriction as to the
number of intermediate bosons $\sigma $. We discuss how to build a
systematic approximation to the kernel of the integral equation for the two
body Green function as a function of the number of particles in the
intermediate Fock state and the power in the coupling constant. A consistent
truncation can be carried out and in the lowest order, this brings to the
Weinberg's equation for the bound state \cite{10}.

This diagram brings about a new feature which was not present in the
previous diagram just considered. Here we come across not only with all
those diagrams that involve the propagation of information to future times,
but note one particular diagram that bears pair production, i.e., there is
one diagram which has intrinsic propagation of information to the past, thus
mingling the two sectorized Fock spaces of solutions. This diagram, as far
as we known, has not being considered in the literature before.

Explicitly this correction to the propagator is written down in terms of the
one boson propagators and is given by the following equation%
\begin{eqnarray}
\Delta S_{\times }(x^{+}) &=&(ig)^{4}\int d\overline{x}_{1}^{+}d\overline{x}%
_{2}^{+}d\overline{x}_{3}^{+}d\overline{x}_{4}^{+}S_{3}(x^{+}-\overline{x}%
_{2}^{+})S_{\sigma }(\overline{x}_{2}^{+}-\overline{x}_{3}^{+})S_{2}(%
\overline{x}_{2}^{+}-\overline{x}_{1}^{+})S_{6}(x^{+}-\overline{x}%
_{4}^{+})\times  \notag \\
&&S_{\sigma }(\overline{x}_{4}^{+}-\overline{x}_{1}^{+})S_{5}(\overline{x}%
_{4}^{+}-\overline{x}_{3}^{+})\times S_{1}(\overline{x}_{1}^{+})S_{4}(%
\overline{x}_{3}^{+}),  \label{1c}
\end{eqnarray}%
and after Fourier transform we have

\begin{eqnarray}
\Delta \widetilde{S}_{\times }(K^{-}) &=&\frac{\left( ig\right) ^{4}i^{8}}{%
2^{3}\left( 2\pi \right) ^{3}}\int dk^{-}dp^{-}dq^{-}\frac{%
dp^{+}d^{2}p_{\perp }}{k^{+}p^{+}(k-p)^{+}(p-q)^{+}(K-k-q+p)^{+}}\times 
\notag \\
&&\frac{1}{\left( K-q\right) ^{+}\left( q-p\right) ^{+}\left( k-p\right) ^{+}%
}\times  \notag \\
&&\frac{1}{\left( k^{-}-\frac{k_{\perp }^{2}+m^{2}-i\varepsilon }{k^{+}}%
\right) }\frac{1}{\left( p^{-}-\frac{k_{\perp }^{2}+m^{2}-i\varepsilon }{%
p^{+}}\right) }\times  \notag \\
&&\frac{1}{\left( q^{-}-\frac{k_{\perp }^{2}+m^{2}-i\varepsilon }{q^{+}}%
\right) }\frac{1}{\left( K^{-}-k^{-}-\frac{\left( K-k\right) _{\perp
}^{2}+m^{2}-i\varepsilon }{\left( K-k\right) ^{+}}\right) }  \notag \\
&&\frac{1}{\left( K^{-}-q^{-}-\frac{\left( K-q\right) _{\perp
}^{2}+m^{2}-i\varepsilon }{\left( K-q\right) ^{+}}\right) }\frac{1}{\left(
k^{-}-p^{-}-\frac{\left( k-p\right) _{\perp }^{2}+m_{\sigma
}^{2}-i\varepsilon }{\left( k-p\right) ^{+}}\right) }  \notag \\
&&\frac{1}{\left( K^{-}-k^{-}-q^{-}+p^{-}-\frac{\left( K-k-q+p\right)
_{\perp }^{2}+m^{2}-i\varepsilon }{\left( K-k-q+p\right) ^{+}}\right) } 
\notag \\
&&\frac{1}{\left( q^{-}-p^{-}-\frac{\left( q-p\right) _{\perp
}^{2}+m_{\sigma }^{2}-i\varepsilon }{\left( q-p\right) ^{+}}\right) },
\label{2c}
\end{eqnarray}

For $K^{+}>0$, the regions of integration in $p^{+}$ which define the
position of poles in the complex $p^{-}$ are:

a) $0<q^{+}<p^{+}<k^{+}<K^{+}$

b) $0<k^{+}<p^{+}<q^{+}<K^{+}$

c) $0<p^{+}<q^{+}<k^{+}<K^{+}$

d) $0<p^{+}<k^{+}<q^{+}<K^{+}$

e) $0<k^{+}<q^{+}<p^{+}<K^{+}$

f) $0<q^{+}<k^{+}<p^{+}<K^{+}$

For regions \textquotedblleft $c$ \textquotedblright\ and \textquotedblleft $%
d$ \textquotedblright\ we use the method of partial fracioning twice to
integrate in $p^{-}$ \cite{14}; for \textquotedblleft $a$ \textquotedblright
, \textquotedblleft $b$ \textquotedblright , \textquotedblleft $e$
\textquotedblright\ and \textquotedblleft $f$ \textquotedblright\ this is
not necessary; for regions \textquotedblleft $e$ \textquotedblright\ and
\textquotedblleft $f$ \textquotedblright\ the integration in $p^{-}$
vanishes. \ The diagrams for each of these regions are shown in figure (\ref%
{f4.2}).

\FRAME{fpFU}{4.1891in}{3.4584in}{0pt}{\Qcb{Crossed ladder diagrams time
ordered in $x^{+}>0$.}}{\Qlb{f4.2}}{fg7r4.gif}{\special{language "Scientific
Word";type "GRAPHIC";maintain-aspect-ratio TRUE;display "USEDEF";valid_file
"F";width 4.1891in;height 3.4584in;depth 0pt;original-width
4.625in;original-height 3.8121in;cropleft "0";croptop "1";cropright
"1";cropbottom "0";filename 'E:/unifei/artigo2005/fg7r4.GIF';file-properties
"XNPEU";}}

After performing the analytic integrations in $k^{-},p^{-}$ and $q^{-}$ we
have 
\begin{equation}
\Delta \widetilde{S}_{\times }(K^{-})=\Delta \widetilde{S}_{\times
}^{a}(K^{-})+\Delta \widetilde{S}_{\times }^{b}(K^{-})+\Delta \widetilde{S}%
_{\times }^{c}(K^{-})+\Delta \widetilde{S}_{\times }^{d}(K^{-})  \label{3c}
\end{equation}%
where 
\begin{eqnarray}
\Delta \widetilde{S}_{\times }^{a}(K^{-}) &=&(ig)^{4}\int \frac{%
idp^{+}d^{2}p_{\perp }\theta (k^{+}-p^{+})\theta (p^{+}-q^{+})}{%
2k^{+}(K-k)^{+}\left( K^{-}-\frac{k_{\perp }^{2}+m^{2}-i\varepsilon }{k^{+}}-%
\frac{(K-k)_{\perp }^{2}+m^{2}-i\varepsilon }{(K-k)^{+}}\right) }  \notag \\
&&\frac{1}{2p^{+}(k-p)^{+}(p-q)^{+}(K-k-q+p)^{+}}\times  \notag \\
&&\frac{i}{\left( K^{-}-\frac{p_{\perp }^{2}+m^{2}-i\varepsilon }{p^{+}}-%
\frac{(K-k)_{\perp }^{2}+m^{2}-i\varepsilon }{(K-k)^{+}}-\frac{(k-p)_{\perp
}^{2}+m^{2}-i\varepsilon }{(k-p)^{+}}\right) }\times  \notag \\
&&\frac{i}{\left( K^{-}-\frac{(p-q)_{\perp }^{2}+m_{\sigma
}^{2}-i\varepsilon }{(p-q)^{+}}-\frac{(k-p)_{\perp }^{2}+m_{\sigma
}^{2}-i\varepsilon }{(k-p)^{+}}-\frac{(K-k)_{\perp }^{2}+m^{2}-i\varepsilon 
}{(K-k)^{+}}-\frac{q_{\perp }^{2}+m^{2}-i\varepsilon }{q^{+}}\right) }\times
\notag \\
&&\frac{i}{\left( K^{-}-\frac{q_{\perp }^{2}+m^{2}-i\varepsilon }{q^{+}}-%
\frac{(k-p)_{\perp }^{2}+m_{\sigma }^{2}-i\varepsilon }{(k-p)^{+}}-\frac{%
(K-k-q+p)_{\perp }^{2}+m^{2}-i\varepsilon }{(K-k-q+p)^{+}}\right) }  \notag
\\
&&\frac{i}{2q^{+}(K-q)^{+}\left( K^{-}-\frac{q_{\perp
}^{2}+m^{2}-i\varepsilon }{q^{+}}-\frac{(K-q)_{\perp
}^{2}+m^{2}-i\varepsilon }{(K-q)^{+}}\right) },  \label{4c}
\end{eqnarray}%
and 
\begin{equation}
\Delta \widetilde{S}_{\times }^{b}(K^{-})=\Delta \widetilde{S}_{\times
}^{a}(K^{-})[k\leftrightarrow q],  \label{5c}
\end{equation}%
shown in figures (\ref{f4.2}a) and (\ref{f4.2}b) respectively.

Regions \textquotedblright $c$\textquotedblright\ and \textquotedblright $d$%
\textquotedblright\ contribute to the propagator correction as 
\begin{eqnarray}
\Delta \widetilde{S}_{\times }^{c}(K^{-}) &=&(ig)^{4}\int \frac{%
idp^{+}d^{2}p_{\perp }\theta (q^{+}-p^{+})\theta (k^{+}-p^{+})}{%
2k^{+}(K-k)^{+}\left( K^{-}-\frac{k_{\perp }^{2}+m^{2}-i\varepsilon }{k^{+}}-%
\frac{(K-k)_{\perp }^{2}+m^{2}-i\varepsilon }{(K-k)^{+}}\right) }  \notag \\
&&\frac{1}{2p^{+}(k-p)^{+}(p-q)^{+}(K-k-q+p)^{+}}\times  \notag \\
&&\frac{i}{\left( K^{-}-\frac{p_{\perp }^{2}+m^{2}-i\varepsilon }{p^{+}}-%
\frac{(q-p)_{\perp }^{2}+m_{\sigma }^{2}-i\varepsilon }{(q-p)^{+}}-\frac{%
(K-k-q+p)_{\perp }^{2}+m^{2}-i\varepsilon }{(K-k-q+p)^{+}}-\frac{%
(k-p)_{\perp }^{2}+m_{\sigma }^{2}-i\varepsilon }{(k-p)^{+}}\right) }\times 
\notag \\
&&\widetilde{S^{\prime }}_{\times }^{c}\frac{i}{2q^{+}(K-q)^{+}\left( K^{-}-%
\frac{q_{\perp }^{2}+m^{2}-i\varepsilon }{q^{+}}-\frac{(K-q)_{\perp
}^{2}+m^{2}-i\varepsilon }{(K-q)^{+}}\right) },  \label{6c}
\end{eqnarray}%
where

\begin{eqnarray}
\widetilde{S^{\prime }}_{\times }^{c} &=&\frac{i}{\left( K^{-}-\frac{%
q_{\perp }^{2}+m^{2}-i\varepsilon }{q^{+}}-\frac{(k-p)_{\perp
}^{2}+m_{\sigma }^{2}-i\varepsilon }{(k-p)^{+}}-\frac{(K-k-q+p)_{\perp
}^{2}+m^{2}-i\varepsilon }{(K-k-q+p)^{+}}\right) }\times  \notag \\
&&\frac{i}{\left( K^{-}-\frac{k_{\perp }^{2}+m^{2}-i\varepsilon }{k^{+}}-%
\frac{(K-k-q+p)_{\perp }^{2}+m_{\sigma }^{2}-i\varepsilon }{(K-k-q+p)^{+}}-%
\frac{(q-p)_{\perp }^{2}+m^{2}-i\varepsilon }{(q-p)^{+}}\right) }\times 
\notag \\
&&\frac{i}{\left( K^{-}-\frac{k_{\perp }^{2}+m^{2}-i\varepsilon }{k^{+}}-%
\frac{(q-p)_{\perp }^{2}+m_{\sigma }^{2}-i\varepsilon }{(q-p)^{+}}-\frac{%
(K-k-q+p)_{\perp }^{2}+m^{2}-i\varepsilon }{(K-k-q+p)^{+}}\right) }\times 
\notag \\
&&\frac{i}{\left( K^{-}-\frac{p_{\perp }^{2}+m^{2}-i\varepsilon }{p^{+}}-%
\frac{(q-p)_{\perp }^{2}+m_{\sigma }^{2}-i\varepsilon }{(q-p)^{+}}-\frac{%
(K-q)_{\perp }^{2}+m^{2}-i\varepsilon }{(K-q)^{+}}\right) }.  \label{7c}
\end{eqnarray}%
The perturbative correction to the two boson propagator in Eq.(\ref{6c}) is
represented by diagrams indicated in figure (\ref{f4.2}c). The correction
represented by diagrams in figure (\ref{f4.2}d) is given by: 
\begin{equation}
\Delta \widetilde{S}_{\times }^{d}(K^{-})=\Delta \widetilde{S}_{\times
}^{c}(K^{-})\left[ k\leftrightarrow K-k,p\leftrightarrow
K-k-q+p,q\leftrightarrow K-q\right] .  \label{8c}
\end{equation}

\section{Antiparticle contribution}

Next we deduce the antiparticle contribution to the crossed ladder diagram.
This contribution happens for $p^{+}<0$ and $K^{+}-k^{+}-q^{+}+p^{+}<0$. Let
us analyse the first case, $p^{+}<0.$

The region for the $"+"$ component momentum that allows the pole positioned
in both hemispheres of the complex $p^{-}$ plane, and therefore giving
non-vanishing residue, are $-k^{+}<p^{+}<0$ and $\left\vert p^{+}\right\vert
+k^{+}+q^{+}<K^{+}$. So, the result for the momentum integration in $"-"$
component for $0<q^{+}<K^{+}$ and $0<k^{+}<K^{+}$ which correspond to the
non-vanishing results for integrations in $k^{-}$ and $q^{-}$ for $%
-k^{+}<p^{+}<0$, is given by the diagram depicted in figure (\ref{f12}) and
the result is given in:

\FRAME{fpFU}{2.8556in}{1.0542in}{0pt}{\Qcb{Pair creation process
contributing to the crossed ladder diagram.}}{\Qlb{f12}}{f12a.gif}{\special%
{language "Scientific Word";type "GRAPHIC";maintain-aspect-ratio
TRUE;display "USEDEF";valid_file "F";width 2.8556in;height 1.0542in;depth
0pt;original-width 2.8124in;original-height 1.0205in;cropleft "0";croptop
"1";cropright "1";cropbottom "0";filename
'E:/unifei/artigo2005/f12a.GIF';file-properties "XNPEU";}} 
\begin{eqnarray}
\Delta \widetilde{S}_{\times }^{ant}(K^{-}) &=&\frac{(ig)^{4}}{8}\int
dp^{+}d^{2}p_{\perp }\frac{\theta (k^{+}+\left\vert p^{+}\right\vert )\theta
(q^{+}+\left\vert p^{+}\right\vert )}{k^{+}\left\vert p^{+}\right\vert
q^{+}(\left\vert p\right\vert +k)^{+}(q+\left\vert p\right\vert )^{+}}\times
\notag \\
&&\frac{\theta (K^{+}-\left\vert p^{+}\right\vert -q^{+}-k^{+})}{\left(
K-k\right) ^{+}\left( K-q\right) ^{+}(K-\left\vert p\right\vert -q-k)^{+}}%
\times  \notag \\
&&\frac{i}{(K^{-}-K_{0}^{-})}\frac{i}{(K^{-}-Q_{0}^{-})}\frac{i}{%
(K^{-}-T^{-})}\times  \notag \\
&&\frac{i}{(K^{-}-J_{a}^{-})}\frac{i}{(K^{-}-T^{\prime -})}+  \label{9c} \\
+[k &\leftrightarrow &q]+[k\rightarrow K-k,q\rightarrow K-q]+[k\rightarrow
K-q,q\rightarrow K-k],  \notag
\end{eqnarray}%
where 
\begin{eqnarray}
T^{-} &=&\frac{(p-q)_{\bot }^{2}+m_{\sigma }^{2}}{q^{+}+\left\vert
p^{+}\right\vert }+\frac{(K+p-q-k)_{\bot }^{2}+m^{2}}{K^{+}-k^{+}-q^{+}-%
\left\vert p^{+}\right\vert }+  \notag \\
&&+\frac{k_{\bot }^{2}+m^{2}}{k^{+}},  \notag \\
J_{a}^{-} &=&\frac{q_{\bot }^{2}+m^{2}}{q^{+}}+\frac{(K-k-q+p)_{\bot
}^{2}+m^{2}}{K^{+}-k^{+}-q^{+}-\left\vert p^{+}\right\vert }+  \notag \\
&&+\frac{p_{\bot }^{2}+m^{2}}{\left\vert p^{+}\right\vert }+\frac{k_{\bot
}^{2}+m^{2}}{k^{+}},  \notag \\
T^{\prime -} &=&\frac{(q-p)_{\bot }^{2}+m_{\sigma }^{2}}{q^{+}+\left\vert
p^{+}\right\vert }+\frac{(K-k-w+p)_{\bot }^{2}+m^{2}}{K^{+}-k^{+}-q^{+}-%
\left\vert p^{+}\right\vert }+  \notag \\
&&+\frac{q_{\bot }^{2}+m^{2}}{q^{+}}.  \label{10c}
\end{eqnarray}%
\qquad

The four body propagator, $J_{a}^{-},$\ (subindex $a$ for antiparticle) has
a propagation to past in the null-plane of an antiparticle with $p^{+}<0$.
At instant $\overline{x}_{2}^{+}>0$ the pair particle-antiparticle is
produced by the $\sigma $ intermediate boson, then the antiparticle
encounters a particle of momentum $k^{+}>0$, and is anihilated and the
production of a $\sigma $ boson with momentum $\left\vert p^{+}\right\vert
+k^{+}>0$ which continues to propagate into the future of the null-plane.

\section{Hierarchical equations}

In general, the Green function in the light front for a system of two bodies
could be obtained from the solution for the covariant Bethe-Salpeter
equation which has all the two body irreducible diagrams in the kernel and
self-energy corrections to the intermediate propagators of the $\phi _{1}$
and $\phi _{2}$ bosons. We can easily obtain the two boson Green function in
the light front, without including self-energy corrections to the
intermediate bosons, that is, closed loops for the bosons $\phi _{1}$ and $%
\phi _{2}$ and crossed diagrams, as a solution to the following set of
hierarchical equations: 
\begin{eqnarray}
G^{(2)}(K^{-})
&=&G_{0}^{(2)}(K^{-})+G_{0}^{(2)}(K^{-})VG^{(3)}(K^{-})VG^{(2)}(K^{-})\ , 
\notag \\
G^{(3)}(K^{-})
&=&G_{0}^{(3)}(K^{-})+G_{0}^{(3)}(K^{-})VG^{(4)}(K^{-})VG^{(3)}(K^{-})\ , 
\notag \\
G^{(4)}(K^{-})
&=&G_{0}^{(4)}(K^{-})+G_{0}^{(4)}(K^{-})VG^{(5)}(K^{-})VG^{(4)}(K^{-})\ , 
\notag \\
&&\ .\ .\ .  \notag \\
G^{(N)}(K^{-})
&=&G_{0}^{(N)}(K^{-})+G_{0}^{(N)}(K^{-})VG^{(N+1)}(K^{-})VG^{(N)}(K^{-})\ , 
\notag \\
&&\ .\ .\ .  \label{3.26}
\end{eqnarray}

The set of equations above, (\ref{3.26}), include, in particular, the
\textquotedblleft ladder\textquotedblright\ approximation for the covariant
Bethe-Salpeter equation. The hierarchical equations, (\ref{3.26}),
correspond to a truncation in the Fock space in the light front, so that
only states with two particles $\phi _{1}$ and $\phi _{2}$ without
restriction in the number of $\sigma $ bosons are permitted in the
intermediate states. The free propagation of these states is represented by
the Green function $G_{0}^{(N)}(K^{-})$, where the number of $\sigma $
bosons is $N-2$. The equations (\ref{3.26}) do not, however, include the
totality of crossed \textquotedblleft ladder\textquotedblright\ diagrams.
For example, the intermediate propagation in the light front of a state of
one $\phi _{1}$ boson, two $\phi _{2}$ bosons and one $\phi _{2}$ antiboson
(four body Fock state) are not included in the proposed hierarchical
equations. In order to get the two body propagator in the light front time
in the \textquotedblleft ladder\textquotedblright\ approximation, we shall
restrict ourselves to the hierarchy of equations (\ref{3.26}).

\section{Bethe-Salpeter equation in the light front}

In order to build the integral equation for the vertex of the bound state of
two interacting bosons in the light front, we need to introduce the concept
of ``vertex'' associated with a wave function of a bound state. To make our
discussion more didatic, we introduce the concept of ``vertex'' through Schr%
\"{o}dinger's equation for the two body bound state, which for the center of
mass is given by 
\begin{equation}
\left( \frac{p^{2}}{2m}+V\right) \left| \Psi _{B}\right\rangle =-E_{B}\left|
\Psi _{B}\right\rangle ,  \label{5.0}
\end{equation}
where $\overrightarrow{p}$ is the relative momentum of the two particles
with same mass $m$, $V$ is the potential and $-E_{B}$ is the bounding
energy. Solving (\ref{5.0}), we have that 
\begin{equation}
\left| \Psi _{B}\right\rangle =\frac{1}{\left( -E_{B}-\frac{p^{2}}{m}\right) 
}V\left| \Psi _{B}\right\rangle ,  \label{5.01}
\end{equation}
where the vertex is given by $V\left| \Psi _{B}\right\rangle =\left| \Gamma
_{B}\right\rangle $.

Our interest is to define the vertex of the bound state of two bosons using
it later on in the hierarchical equations of the Green function (\ref{3.26}%
), from which we get the non perturbative equation for the propagator of two
bosons: 
\begin{equation}
G^{\left( 2\right) }=G_{0}^{\left( 2\right) }+G_{0}^{\left( 2\right)
}VG^{\left( 3\right) }VG^{\left( 2\right) },  \label{5.1}
\end{equation}
which allows, in principle, the appearance of bound states.

Close to the region of bound state energy the Green function has a pole: 
\begin{equation}
\lim_{K^{-}\rightarrow K_{B}^{-}}G^{\left( 2\right) }(K^{-})=\frac{\left|
\psi _{B}\right\rangle \left\langle \psi _{B}\right| }{K^{-}-K_{B}^{-}},
\label{5.2}
\end{equation}
where $\left| \psi _{B}\right\rangle $ is the wave function of the bound
state.

So, introducing (\ref{5.2}) into (\ref{5.1}) and taking the limit $%
K^{-}\rightarrow K_{B}^{-}$, we have 
\begin{equation}
|\Psi _{B}>=G_{0}^{(2)}(K_{B}^{-})VG^{(3)}(K_{B}^{-})V|\Psi _{B}>,
\label{5.3}
\end{equation}
with the kernel defined by the hierarchical equations (\ref{3.26}).

This can also be written as an eigenvalue equation for a mass squared
operator 
\begin{equation}
\left[ M_{0}^{2}+K^{+}VG^{(3)}(K_{B}^{-})V\right] |\Psi
_{B}>=(M_{2})^{2}|\Psi _{B}>,  \label{5.4}
\end{equation}%
where $(M_{2})^{2}=K^{+}K_{B}^{-}-K_{\perp }^{2},$ $%
M_{0}^{2}=K^{+}K_{(2)on}^{-}-K_{\perp }^{2}$ and $K_{(2)on}^{-}$ is defined
in

\begin{equation*}
K_{(2)on}^{-}=\frac{k_{1\perp }^{2}+m^{2}}{k_{1}^{+}}+\frac{k_{2\perp
}^{2}+m^{2}}{k_{2}^{+}}
\end{equation*}

Equation (\ref{5.3}) is the homogeneous Bethe-Salpeter equation projected
onto the light front, and the vertex is defined by, 
\begin{equation}
\left| \Gamma _{B}\right\rangle =\left( G_{0}^{(2)}(K_{B}^{-})\right)
^{-1}|\Psi _{B}>.  \label{5.5}
\end{equation}

\section{Ladder approximation in $O(g^{2})$}

To obtain the homogeneous integral equation for the vertex up to $g^{2}$
order, multiply (\ref{5.3}) by $\left( G_{0}^{(2)}\right) ^{-1}$ on both
sides and using the property $G_{0}^{(2)}\left( G_{0}^{(2)}\right) ^{-1}=1$,
we have: 
\begin{equation}
\left| \Gamma _{B}\right\rangle =VG_{0}^{(3)}VG_{0}^{(2)}\left| \Gamma
_{B}\right\rangle .  \label{5.6}
\end{equation}

In the basis of kinematical momenta and defining the momentum fractions $x=%
\frac{k^{+}}{K^{+}}$, and $y=\frac{q^{+}}{K^{+}}$, we have 
\begin{equation}
\Gamma _{B}(\overrightarrow{q}_{\perp },y)=\int \frac{dxd^{2}k_{\perp }}{%
x(1-x)}\frac{\mathit{K}^{(3)}(\overrightarrow{q}_{\perp },y;\overrightarrow{k%
}_{\perp },x)}{M_{B}^{2}-M_{0}^{2}}\Gamma _{B}(\overrightarrow{k}_{\perp
},x),  \label{5.7}
\end{equation}
where the free mass of the two boson system, in the center of mass $%
\overrightarrow{K}_{\perp }=0$, is given by 
\begin{equation}
M_{0}^{2}=K^{+}K_{(2)on}^{-}-K_{\perp }^{2}=\frac{k_{\perp }^{2}+m^{2}}{%
x(1-x)},  \label{5.8}
\end{equation}
and 
\begin{eqnarray}
\mathit{K}^{(3)}(\overrightarrow{q}_{\perp },y;\overrightarrow{k}_{\perp
},x) &=&\frac{g^{2}}{16\pi ^{3}}\frac{\theta (x-y)}{\left( x-y\right) }\times
\label{5.9} \\
&&\frac{1}{\left( M_{B}^{2}-\frac{q_{\perp }^{2}+m^{2}}{y}-\frac{k_{\perp
}^{2}+m^{2}}{1-x}-\frac{\left( q-k\right) _{\perp }^{2}+m_{\sigma }^{2}}{x-y}%
\right) }+  \notag \\
&&+\left[ k\leftrightarrow q\right] ,  \notag
\end{eqnarray}
with $M_{B}^{2}=K_{B}^{+}K_{B}^{-}$, in the center of mass where $%
\overrightarrow{K}_{\perp }=0$. We draw attention to the fact that $\left|
\Gamma _{B}\right\rangle $ and $\Gamma _{B}(\overrightarrow{q}_{\perp },y)$
are related to each other by a phase space factor, such that $\Gamma _{B}(%
\overrightarrow{q}_{\perp },y)=\sqrt{q^{+}(K^{+}-q^{+})}\left\langle 
\overrightarrow{q}_{\perp },q^{+}\right. \left| \Gamma _{B}\right\rangle $.

Equation (\ref{5.7}) is the Bethe-Salpeter equatin in the null plane in the
``ladder'' approximation with the interaction calculated in second order in
the coupling constant. This is also the same as the equation obtained by
Weinberg \cite{10}.

\section{Ladder approximation in $O(g^{4})$}

We build the homogeneous integral equation for the vertex of the bound state
in the light front starting from (\ref{3.26}) which has the approximate
kernel to order $g^{4}$, whose non perturbative solution resulst in the
Green function of two bosons: 
\begin{equation*}
G_{g^{4}}^{(2)}(K^{-})=G_{0}^{(2)}(K^{-})+G_{0}^{(2)}(K^{-})VG^{(3)}(K^{-})VG_{g^{4}}^{(2)}(K^{-}),
\end{equation*}%
where 
\begin{eqnarray*}
G^{(3)}(K^{-}) &\cong &G_{0}^{(3)}(K^{-})+\Delta G_{g^{2}}^{(3)}(K^{-})= \\
&=&G_{0}^{(3)}(K^{-})+G_{0}^{(3)}(K^{-})VG_{0}^{(4)}(K^{-})VG_{0}^{(3)}(K^{-}).
\end{eqnarray*}

Truncate now $G^{(3)}$ to order $g^{2}$ and $G^{(4)}$ up to $g^{0}$,
including in this way, the intermediate propagations of three and four
bodies in the kernel of the integral equation for $G_{g^{4}}^{(2)}$.

The integral equation for the vertex of the bound state through the fourth
order is buit in an analogous form as done previously, where the vertex is
given now by 
\begin{equation}
\left| \Gamma _{B}\right\rangle =V\left( G_{0}^{(3)}(K^{-})+\Delta
G_{g^{2}}^{(3)}(K^{-})\right) VG_{0}^{(2)}\left| \Gamma _{B}\right\rangle .
\label{5.10a}
\end{equation}

We define the incoming particles in the diagram by the fraction of momentum, 
$x=\frac{k^{+}}{K^{+}}$, and $\overrightarrow{k}_{\perp }$, where as the
outgoing one by $y=\frac{q^{+}}{K^{+}}$ and $\overrightarrow{q}_{\perp }$.
The momenta flowing in the loop which generates the intermediate state of
four bodies we define as e a do $z=\frac{p^{+}}{K^{+}}$ and $\overrightarrow{%
p}_{\perp }$, so that we have: 
\begin{eqnarray}
\Gamma _{B}(\overrightarrow{q}_{\perp },y) &=&\int \frac{dxd^{2}k_{\perp }}{%
x(1-x)}\times  \label{5.10} \\
&&\frac{\mathit{K}^{(3)}(\overrightarrow{q}_{\perp },y;\overrightarrow{k}%
_{\perp },x)+\mathit{K}^{(4)}(\overrightarrow{q}_{\perp },y;\overrightarrow{k%
}_{\perp },x)}{M_{B}^{2}-M_{0}^{2}}\Gamma _{B}(k_{\perp },x),  \notag
\end{eqnarray}
with 
\begin{eqnarray}
\mathit{K}^{(4)}(\overrightarrow{q}_{\perp },y;\overrightarrow{k}_{\perp
},x) &=&\left( \frac{g^{2}}{16\pi ^{3}}\right) ^{2}\int \frac{d^{2}p_{\perp
}dz}{z\left( z-x\right) \left( y-z\right) \left( 1-z\right) }  \notag \\
&&\frac{\theta (y-z)\theta (z-x)}{\left( M_{B}^{2}-\frac{k_{\perp }^{2}+m^{2}%
}{x}-\frac{p_{\perp }^{2}+m^{2}}{1-z}-\frac{\left( p-k\right) _{\perp
}^{2}+m_{\sigma }^{2}}{z-x}\right) }  \notag \\
&&\frac{1}{\left( M_{B}^{2}-\frac{p_{\perp }^{2}+m^{2}}{z}-\frac{q_{\perp
}^{2}+m^{2}}{1-y}-\frac{\left( q-p\right) _{\perp }^{2}+m_{\sigma }^{2}}{y-z}%
\right) }  \notag \\
&&\frac{1}{\left( M_{B}^{2}-\frac{k_{\perp }^{2}+m^{2}}{x}-\frac{q_{\perp
}^{2}+m^{2}}{1-y}-\frac{\left( q-p\right) _{\perp }^{2}+m_{\sigma }^{2}}{y-z}%
+\frac{\left( p-k\right) _{\perp }^{2}+m_{\sigma }^{2}}{z-x}\right) }  \notag
\\
&&+\left[ x\leftrightarrow y,k_{\perp }\leftrightarrow q_{\perp }\right] .
\label{5.11}
\end{eqnarray}

The vertex $\left| \Gamma _{B}\right\rangle $ and $\Gamma _{B}( 
\overrightarrow{q}_{\perp },y)$ are related like: 
\begin{equation*}
\left\langle \overrightarrow{q}_{\perp },q^{+}\right. \left| \Gamma
_{B}\right\rangle =\sqrt{q^{+}(K^{+}-q^{+})}\Gamma _{B}(\overrightarrow{q}
_{\perp },y)
\end{equation*}

The equation (\ref{5.10}) is the Bethe-Salpeter equation in the light front
with the kernel to the fourth order in the coupling constant in the
\textquotedblleft ladder\textquotedblright\ approximation. It is to be noted
that we were able to construct the Bethe-Salpeter equation in the
\textquotedblleft ladder\textquotedblright\ approximation from the Green
function in the light front, and if we need to go to higher orders in $g$ it
is only necessary to add terms that include the propagation of $N$ bodies in
the non perturbative hierarchical equations (\ref{3.26}).

\section{Cross ladder}

The construction of the integral equation for the vertex follows the same
procedure done for the diagram stairway, however we have to add in Eq.(\ref%
{5.10a}) the function of Green for the crossed diagram

\begin{eqnarray}
|\Psi _{B} &>&=G_{0}^{(2)}(K_{B}^{-})V\{G_{0}^{(3)}(K_{B}^{-})+  \label{1b}
\\
&&G_{0}^{(3)}(K_{B}^{-})V[G^{(4)}(K_{B}^{-})+G_{x}^{(4)}(K_{B}^{-})+G_{x}^{(4)anti}(K_{B}^{-})]VG_{0}^{(3)}(K_{B}^{-})\}V\left\vert \Psi _{B}\right\rangle
\notag
\end{eqnarray}

The vertex for the linked state satisfies the following homogeneous integral
equation now

\begin{eqnarray}
\Gamma _{B}(\overrightarrow{q}_{\perp },y) &=&\int \frac{dxd^{2}k_{\perp }}{%
x(1-x)}\frac{\Gamma _{B}(\overrightarrow{k}_{\perp },x)}{M_{B}^{2}-M_{o}^{2}}%
\times  \label{2b} \\
&&\left[ \mathit{K}^{(3)}(q_{\perp },y;\overrightarrow{k}_{\perp },x)+%
\mathit{K}^{(4)}(\overrightarrow{q}_{\perp },y;\overrightarrow{k}_{\perp
},x)+\right.  \notag \\
&&\left. \mathit{K}_{\times }^{(4)}(\overrightarrow{q}_{\perp },y;%
\overrightarrow{k}_{\perp },x)+\mathit{K}_{\times }^{(4)anti}(%
\overrightarrow{q}_{\perp },y;\overrightarrow{k}_{\perp },x)\right] ,  \notag
\end{eqnarray}%
where\ 
\begin{equation}
\mathit{K}_{\times }^{(4)}=\left( \frac{g^{2}}{16\pi ^{3}}\right) ^{2}\int 
\frac{d^{2}p_{\perp }dz\theta (1-z)}{z\left( x-z\right) \left(
1-z-x-y\right) }\left( \mathit{K}_{\times }^{\prime }+\mathit{K}_{\times
}^{^{\prime \prime }}\right) ,  \label{3b}
\end{equation}%
with 
\begin{eqnarray}
\mathit{K}_{\times }^{\prime } &=&\frac{\theta (z-y)\theta (x-z)}{(z-y)\left[
M_{B}^{2}-\frac{p_{\perp }^{2}+m^{2}-i\varepsilon }{z}-\frac{k_{\perp
}^{2}+m^{2}-i\varepsilon }{1-x}-\frac{(k-p)_{\perp }^{2}+m_{\sigma
}^{2}-i\varepsilon }{x-z}\right] }\times  \notag \\
&&\frac{1}{M_{B}^{2}-\frac{(p-q)_{\perp }^{2}+m_{\sigma }^{2}-i\varepsilon }{%
z-y}-\frac{(k-p)_{\perp }^{2}+m_{\sigma }^{2}-i\varepsilon }{x-z}-\frac{%
k_{\perp }^{2}+m^{2}-i\varepsilon }{1-x}-\frac{q_{\perp
}^{2}+m^{2}-i\varepsilon }{y}}\times  \notag \\
&&\frac{1}{M_{B}^{2}-\frac{(k-p)_{\perp }^{2}+m_{\sigma }^{2}-i\varepsilon }{%
x-z}-\frac{(p-k-q)_{\perp }^{2}+m^{2}-i\varepsilon }{1-z-x-y}-\frac{q_{\perp
}^{2}+m_{\sigma }^{2}-i\varepsilon }{y}}+  \notag \\
&&\left[ x\leftrightarrow y,k_{\perp }\leftrightarrow q_{\perp }\right] ,
\label{4b}
\end{eqnarray}%
and 
\begin{eqnarray}
\mathit{K}_{\times }^{^{\prime \prime }} &=&\frac{\theta (y-z)\theta (x-z)}{%
(y-z)}\times  \notag \\
&&\frac{1}{M_{B}^{2}-\frac{(k-p)_{\perp }^{2}+m_{\sigma }^{2}-i\varepsilon }{%
x-z}-\frac{(1-p-k-q)_{\perp }^{2}+m^{2}-i\varepsilon }{1-z-x-y}-\frac{%
p_{\perp }^{2}+m^{2}-i\varepsilon }{z}-\frac{(q-p)_{\perp }^{2}+m_{\sigma
}^{2}-i\varepsilon }{y-z}}\times  \notag \\
&&\frac{1}{M_{B}^{2}-\frac{(q-p)_{\perp }^{2}+m_{\sigma }^{2}-i\varepsilon }{%
y-z}-\frac{(1-p-k-q)_{\perp }^{2}+m^{2}-i\varepsilon }{1-z-x-y}-\frac{%
k_{\perp }^{2}+m^{2}-i\varepsilon }{x}}\times  \notag \\
&&\left\{ \frac{1}{M_{B}^{2}-\frac{(k-p)_{\perp }^{2}+m_{\sigma
}^{2}-i\varepsilon }{x-z}-\frac{(1-p-k-q)_{\perp }^{2}+m^{2}-i\varepsilon }{%
1-z-x-y}-\frac{q_{\perp }^{2}+m_{\sigma }^{2}-i\varepsilon }{y}}+\right. 
\notag \\
&&\left. \frac{1}{M_{B}^{2}-\frac{(q-p)_{\perp }^{2}+m_{\sigma
}^{2}-i\varepsilon }{y-z}-\frac{q_{\perp }^{2}+m^{2}-i\varepsilon }{1-y}-%
\frac{p_{\perp }^{2}+m^{2}-i\varepsilon }{z}}\right\} +  \label{5b} \\
&&\left[ x\leftrightarrow 1-x;z\leftrightarrow 1+z-x-y,p_{\perp
}\leftrightarrow p_{\perp }\leftrightarrow (1+p-k-q)_{\perp
};y\leftrightarrow 1-y\right] .  \notag
\end{eqnarray}

The contribution of the pair particle and anti-particle to the kernel of Eq.(%
\ref{2b}) it is given for:

\begin{eqnarray}
\Gamma _{B}(q_{\perp },y) &=&\int \frac{dxd^{2}k_{\perp }}{x(1-x)}\frac{%
\Gamma _{B}(k_{\perp },x)}{M_{B}^{2}-M_{o}^{2}}\left\{ \mathit{K}%
^{(3)}(q_{\perp },y;k_{\perp },x)+\mathit{K}^{(4)}(q_{\perp },y;k_{\perp
},x)\right. +  \notag \\
&&+\left. \mathit{K}_{\times }^{(4)}(q_{\perp },y;k_{\perp },x)+\mathit{K}%
_{\times }^{(4)ant}(q_{\perp },y;k_{\perp },x)\right\} ,  \label{6b}
\end{eqnarray}%
where the kernel for anti-particle is given for: 
\begin{eqnarray}
\mathit{K}_{\times }^{(4)ant} &=&\left( \frac{g^{2}}{16\pi ^{3}}\right)
^{2}\int \frac{d^{2}k_{\perp }dz\theta (1+z)\theta (-z)}{\left\vert
z\right\vert \left( y+\left\vert z\right\vert \right) \left( 1-\left\vert
z\right\vert -x-y\right) (x+\left\vert z\right\vert )}\times  \notag \\
&&\frac{\theta (y+z)\theta (x+z)}{M^{2}-\frac{w_{\perp
}^{2}+m^{2}-i\varepsilon }{y}-\frac{(K+k-p-w)_{\perp
}^{2}+m^{2}-i\varepsilon }{1-\left\vert z\right\vert -x-y}+\frac{k_{\perp
}^{2}+m^{2}-i\varepsilon }{\left\vert z\right\vert }-\frac{p_{\perp
}^{2}+m_{\sigma }^{2}-i\varepsilon }{x}}\times  \notag \\
&&\frac{1}{M^{2}-\frac{(p-k)_{\perp }^{2}+m_{\sigma }^{2}-i\varepsilon }{%
x+\left\vert z\right\vert }-\frac{(K+k-p-w)_{\perp }^{2}+m^{2}-i\varepsilon 
}{1-\left\vert z\right\vert -x-y}-\frac{p_{\perp }^{2}+m_{\sigma
}^{2}-i\varepsilon }{x}}\times  \notag \\
&&\frac{1}{M^{2}-\frac{(w-k)_{\perp }^{2}+m_{\sigma }^{2}-i\varepsilon }{%
y+\left\vert z\right\vert }-\frac{(K+k-p-w)_{\perp }^{2}+m^{2}-i\varepsilon 
}{1-\left\vert z\right\vert -x-y}-\frac{w_{\perp }^{2}+m_{\sigma
}^{2}-i\varepsilon }{y}}.  \label{7b}
\end{eqnarray}

\bigskip

The equation (\ref{2b}) it is the equation of Bethe-Salpeter with the kernel
expanded up to fourth order in g in the null-plan in cross-ladder approach.

\section{Conclusions}

The Bethe-Salpeter equation is constructed via few-body propagation, it has
been demonstrated that the ladder diagram in the light-front gives an
important contribution to the covariance of the process (equal-time three
and four particles)\cite{14}. In our case, we are observing that for the
Bethe-Salpeter construction, besides the equal-time three- and four
particles we also should have a non-vanishing contribution due to pair
production. So we expect and suspect that the crossed ladder diagram in such
a construction will also give a non-vanishing relevant contribution; which
contribution is of the z-diagram type (backward-moving particles).

\textit{Acknowledgments:} J.H.O.Sales (FAPEMIG/cex-1661/05) thanks the
hospitality of Instituto de F\'{\i}sica Te\'{o}rica/UNESP, where part of
this work has been done.A.T.Suzuki thanks the kind hospitality of Physics
Department, North Carolina Staste University and gratefully acknowledges
partial support from CNPq (Bras\'{\i}lia) in the earlier stages of this
work, now superseded by a grant from CAPES (Bras\'{\i}lia).


\begin{thebibliography}{99}
\bibitem{37} M. Sawicki, Phy. Rev. \textbf{D44} (1991)433; Phys. Rev. 
\textbf{D46} (1992) 474.

\bibitem{karmanov} J. Carbonell and V.A. Karmanov , hep-th/0505262.

\bibitem{carbonel} V.A. Karmanov and J. Carbonell, nucl-th/0510051.

\bibitem{13} T.Frederico, J.H.O.Sales, B.V. Carlson and P.U. Sauer, Few-Body
Syst. \textbf{10} (1999)123.

\bibitem{14} J.H.Sales, T. Frederico, B.V. Carlson and P.U. Sauer, Phys.
Rev. \textbf{C} \textbf{61} (2000) 044003.

\bibitem{36} S.J.Brodsky, H.-C.Pauli and S.Pinski, Phys.Rep.\textbf{301}
(1998) 299.

\bibitem{30} L.L.Frankfurt and M.I.Strikman, Nucl. Phys. \textbf{B148}
(1979) 107; Phys. Rep. \textbf{76} (1981)215.

\bibitem{5} L.L. Frankfurt, T. Frederico and M.I. Strikmam, Phys. Rev.%
\textbf{C 48} (1993) 2182. T. Frederico and G.A.Miller, Phy. Rev. \textbf{D45%
} (1992)4207; Phys. Rev. \textbf{D50} (1994) 210.

\bibitem{32} J.Carbonell, B. Desplanques, V. Karmanov and J.F. Mathiot,
Phys.Rep. \textbf{300} (1998) 215.

\bibitem{10} S. Weinberg, Phys. Rev. \textbf{150} (1966)1313.

\bibitem{11} C.-R. Ji and R.J. Furnstahl, Phys. Lett. \textbf{B167}
(1986)11; C.-R. Ji, Phys. Lett. \textbf{B167} (1986)16. C.-R. Ji, Phys.
Lett. \textbf{B322} (1994) 389.

\bibitem{17} S.Glasek, A. Harindranath, S. Pinsky, J.Shigemitsu and K.G.
Wilson, Phys. Rev. \textbf{D47} (1993)1599. R.J. Perry, A. Harindranath and
K.G. Wilson, Phys. Rev. Lett. \textbf{65} (1990) 2959.

\bibitem{41} S.J.Brodsky, C.R.Ji and M.Sawicki, Phys. Rev. \textbf{D32}
(1985)1530.

\bibitem{20} S.D. Drell, D.J. Levy and J.-M. Yan, Phys. Rev. \textbf{D1}
(1970) 1035.

\bibitem{22} J. B.Kogut and D. E.Soper, Phys. Rev. \textbf{D1} (1970) 2901.

\bibitem{18} A. Messiah, \ \textquotedblleft Quantum Mechanics Vol.II
\textquotedblright , North- Holland, Amsterdam, 1962.

\bibitem{1} P.A.M. Dirac, Rev. Mod.Phys. \textbf{21} (1949) 392.

\bibitem{2} J.M.Namyslowski, Progress in Particle and Nuclear Physics 
\textbf{14 }(1985) 49.
\end{thebibliography}
\end{document}